\documentclass[prd,nofootinbib,twocolumn]{revtex4}
\usepackage{graphicx}
\usepackage{bm}
\usepackage{float}
\usepackage{subfigure}
\usepackage{amsmath}
\usepackage{amsfonts}
\usepackage{color}
\usepackage{slashed}
\usepackage{fancyhdr}
\usepackage{ulem}

\def\be{\begin{equation}}
\def\ee{\end{equation}}
\def\bed{\begin{description}}
\def\eed{\end{description}}

\def\bea{\begin{eqnarray}}
\def\eea{\end{eqnarray}}

\def\ba{\begin{array}}
\def\ea{\end{array}}

\def\u1{$U(1)$}
\def\suu1{$SU(2)\times U(1)$}

\begin{document}

\title{Capacitor Discharge and Vacuum Resistance in Massless ${\rm QED}_2$}

\author{Yi-Zen Chu$^1$ and Tanmay Vachaspati$^{1,2}$}
\affiliation{
$^1$CERCA, Department of Physics, Case Western Reserve University,
10900 Euclid Avenue, Cleveland, OH 44106-7079, \\
$^2$Institute for Advanced Study, Princeton, NJ 08540}

\begin{abstract}
\noindent
A charged parallel plate capacitor will create particle-antiparticle
pairs by the Schwinger process and discharge over time. We consider
the full quantum discharge process in 1+1 dimensions including
backreaction, when the electric field interacts with massless
charged fermions. We recover oscillatory features in the electric
field observed in a semiclassical analysis and find that the amplitude
of the oscillations falls off as $t^{-1/2}$ and that \textit{stronger}
coupling implies \textit{slower} decay. Remarkably, Ohm's law applies
to the vacuum and we evaluate the quantum electrical conductivity of
the vacuum to be $e/\sqrt{\pi}$, where $e$ is the fermionic charge.
Similarities and differences with black hole evaporation are mentioned.
\end{abstract}

\maketitle

\section{Introduction}

Following Schwinger's 1951 paper \cite{Schwinger:1951nm}, it is
well-known that quantum effects cause electric fields to produce
electron-positron pairs. The effect can be interpreted as the
tunneling of virtual electron-positron pairs into real particles.
One imagines a virtual $e^+e^-$ pair produced in the vacuum
which is then torn apart by the background electric field, say
within a capacitor, with the
positron accelerated in the direction of the electric field,
and the electron in the opposite direction. A similar effect
has been proposed in de Sitter space where the rapid expansion
of spacetime ``pulls'' particles out of the vacuum.
Hawking radiation from black holes has also been interpreted as
a Schwinger process, though the universally attractive nature
of gravity suggests that there are differences. For example, in
the electromagnetic case, it is clear that once the charges are
created, the positive charge accelerates away from the positively
charged capacitor plate due to electromagnetic repulsion. In the
black hole case, however, there is radiation even though the
black hole attracts all outgoing particles.

The energy for pair creation in an electric field must eventually
come from the energy in the electric field itself. Hence the
electric field has to decay due to the Schwinger process, just
as Hawking radiation is assumed to cause black holes to evaporate.
The problem of electric field decay clearly involves calculation
of the backreaction of the Schwinger process and this is a hard
problem. There have been several attempts to analyze the decay of
the electric field by semiclassical methods, replacing quantum
operators by their expectation values \cite{CooperMottolaEtal}. The results are interesting. For example, a uniform
electric field will not
discharge monotonically but will undergo oscillations.
If the conclusion can be directly transported to the black hole
case, it would imply oscillations of the black hole mass and
not monotonic evaporation. A key difference though is that
electric charges can be positive or negative, whereas the
particles in Hawking radiation can only have positive mass. (See also the recent work \cite{Akhmedov:2009vs}, where backreaction in the context of scalar QED in 3+1 dimensions was taken into account by solving the equations of motion derived from the one loop Euler-Heisenberg effective action.)

In this paper we re-visit the problem of capacitor discharge
due to the Schwinger process, without restricting ourselves to
the semiclassical approximation. We can solve the full quantum
problem but the price we pay is that we are then only able to
treat \textit{massless} fermions and the exponential suppression
of the classic Schwinger process is absent.

We treat the case of massless QED in 1+1 dimensions
\begin{equation}
\mathcal{S}_0 = \int d^2 x \left[
     \bar{\psi} \gamma^\mu (i \partial_\mu + e A_\mu) \psi
        - \frac{1}{4} F_{\mu\nu} F^{\mu\nu} \right]
\end{equation}
The fermions interact with the gauge field by the standard
minimal coupling, and an electric field leads to fermion
pair production. The advantage of 1+1D QED is that it can
be bosonized to yield \cite{Coleman:1976uz,Fujikawa,Naon_PathIntegralBoson},
\begin{equation}
\mathcal{S}'_0 = \int d^2 x \left [
\frac{1}{2} (\partial \phi)^2 + \frac{g}{2} \phi
\epsilon^{\mu\nu} F_{\mu\nu} - \frac{1}{4} F_{\mu\nu} F^{\mu\nu} \right ]
\label{bosonicaction}
\end{equation}
where, $\epsilon_{\mu\nu}$ is the Levi-Civita tensor in 1+1D with
$\epsilon_{01}=1$ and\footnote{{\it Note added}: Previous versions of this paper contained a spurious factor of $2$ in the following formula, i.e., $g \equiv 2e/\sqrt{\pi}$. (We thank Jaume Garriga and Emil Mottola for bringing this to our attention.) We have, accordingly, dropped this factor of $2$ in the electrical conductivity of eq. \eqref{sigmasetup1}, and everywhere in the text referring to it.}
\begin{equation}
g \equiv \frac{e}{\sqrt{\pi}}
\end{equation}
where, without any loss of generality, we can assume $g \ge 0$.
The correspondence between the fermionic and bosonic models is
given by identifying the currents at the quantum operator level
\cite{vonDelft:1998pk}
\begin{equation}
:\bar{\psi} \gamma^\mu \psi : \leftrightarrow
\epsilon^{\mu\nu} \partial_\nu \phi
\end{equation}

The bosonized model is particularly easy to solve because it is
quadratic in fields and hence there are no interactions. It is
sufficient to solve it classically. This simplification only occurs
if the fermions are massless in the original model. If we had included
a mass for the fermions, we would have obtained a sine-Gordon potential
for the scalar field, which is an interacting scalar field theory,
and a quantum treatment of the bosonized model would become necessary.

To analyze the discharge of a capacitor, we would like to set up
an initial electric field that is localized within a finite region
of space and then examine its evolution due to the spontaneous
production of fermion pairs. This leads to a physical
difficulty because the capacitor plates are necessarily external
to the system. This physical difficulty arises even in classical
electromagnetism, where Maxwell's equations are solved but the
boundary conditions are provided externally. The difficulty can
be avoided in gravitational systems, for example, during gravitational
collapse, because the system naturally tends to evolve toward a
black hole. On the other hand, if the gravitational problem is
set up to include an eternal black hole as an ``external'' agent,
a similar issue arises. There is no analog of gravitational
collapse in the electromagnetic situation since similar electric
charges repel. If we were to set up a configuration of $\phi$
field corresponding to separated localized positive and negative
charges, the charge distributions would simply spread out
due to mutual repulsion and then annihilate. Hence an external
capacitor plate is necessary to set up the problem.

Here we will set up the capacitor problem in two ways. In the
first scheme, we fix the boundary condition that there is a
charge $+Q$ at $x=-L/2$ and another charge $-Q$ at $x=+L/2$.
These external charges are taken to be
fixed and are non-dynamical. The Schwinger process then
creates pairs and dissipates the electric field within the
capacitor but cannot ``evaporate'' the charge on the plates.
In this case, we find that the capacitor charges are screened
due to the Schwinger process and the electric field decays
exponentially with distance from a capacitor plate. Our second
strategy to set up the capacitor plates is to introduce an
external potential (two ``bags'') into which we can insert the
charges $+Q$ and $-Q$ as configurations of the $\phi$ field itself.
These charges are now dynamical because the field $\phi$ is
dynamical.  The Schwinger process causes evaporation of the
charge from the bags. In both physical realizations, the
approach to the asymptotic state is not monotonic but oscillatory.
In the first case, the final static state still contains the $\pm Q$
charges but the charges are screened by opposite charges.
In the second case, the final state is not static. Instead it
contains bound states of positive and negative charge densities
that oscillate without dissipation within the bags.

In the following two sections we describe the discharge of a
capacitor in the two physical setups, first with ``external charges''
and second with ``external potential''. We can solve the first
setup analytically, allowing us to obtain explicit expressions
for the late time behavior of the current, electric field,
and energy decay law. The solution of the second setup has only
been obtained numerically. The results of both methods are
summarized in Sec.~\ref{conclusions} and show that the
capacitor discharge is oscillatory, the root-mean-square
current is proportional to the root-mean-square electric
field (Ohm's law), and that the electrical conductivity
of the vacuum is equal to $g=e/\sqrt{\pi}$.

\section{Setup I: External Charges}
\label{setupI}

\begin{figure}
\includegraphics[width=3.5in]{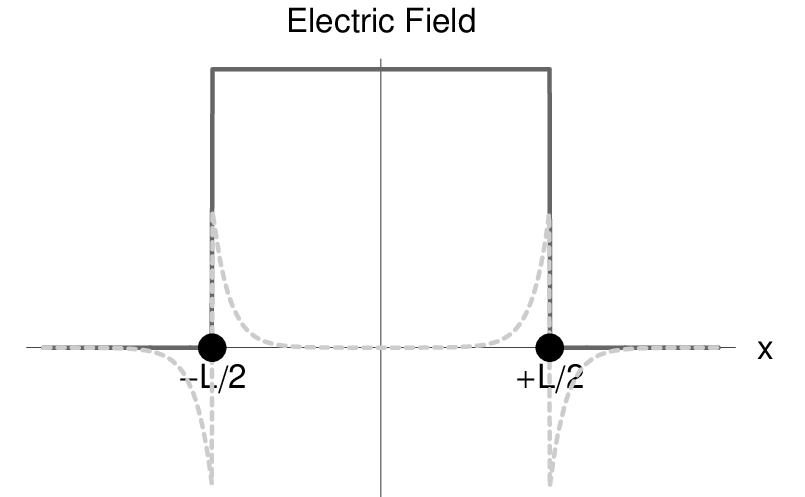}
\caption{Schematic view of setup in section \ref{setupI}. Two
infinitely heavy charges, $+Q$ and $-Q$ are placed at $x=-L/2$
and $x=+L/2$ respectively. The initial electric field is given
by the thick dark line. The expected final electric field is given
by the dashed grey line.}
\label{ParallelPlate_Setup}
\end{figure}

In this section we treat the capacitor as made up of two external
charges $\pm Q$ placed at $x=\mp L/2$ respectively
(see Fig.~\ref{ParallelPlate_Setup}). The electric field due to
these charges satisfies Maxwell's equations and is a non-zero
constant in the region between the plates. The value of the
electric field in the region $-L/2 < x < +L/2$ is $E = Q$.
For $|x| > L/2$, the electric field vanishes. This electric
field configuration, together with $\phi =0$ and
${\dot \phi}\equiv \partial_t\phi =0$, corresponding to no
particles, are the initial conditions whose evolution we
will consider.

The equations of motion follow from (\ref{bosonicaction}).
Including the external charges on the capacitor plates, we get
\begin{eqnarray}
\partial^2 \phi &=& \frac{g}{2} \epsilon^{\mu\nu} F_{\mu\nu}
\label{scalareqn} \\
\partial_\mu F^{\mu\nu} &=&
 j^\nu_\phi + j^\nu_{\rm ext} \equiv j^\nu
\label{maxwelleqn}
\end{eqnarray}
with
\begin{eqnarray}
j^\nu_\phi &=& g \epsilon^{\mu\nu} \partial_\mu \phi    \\
j^\nu_{\rm ext} &=& Q u^\nu (\delta(x-L/2) - \delta(x+L/2))
\end{eqnarray}
and $u^\nu = (1,0)$.

Maxwell's equations (\ref{maxwelleqn}) can be integrated immediately
to get
\begin{equation}
F_{01} = g \phi + \bar{F}
\label{electricfield}
\end{equation}
where
\begin{equation}
{\bar F} = Q (\Theta (x+L/2) - \Theta (x-L/2))
\label{barFsoln}
\end{equation}
Note that we have set the constant of integration to zero so that
the electric field at spatial infinity vanishes.

The charge within some interval $(a,b)$ is given by Gauss' law
\begin{eqnarray}
q &=& F_{01}(t,x=b) - F_{01}(t,x=a) \nonumber \\
  &=& g [\phi(t,x=b) - \phi(t,x=a)]
\end{eqnarray}

Inserting (\ref{electricfield}) into the scalar field equation,
(\ref{scalareqn}), gives
\begin{eqnarray}
(\partial^2 + m^2) \phi &=& - m \bar{F} \nonumber \\
        && \hskip -0.75 cm = - m Q (\Theta (x+L/2) - \Theta (x-L/2))
\label{kleingordoneqn}
\end{eqnarray}
where the effective mass of the scalar field is given by the coupling constant,
\begin{equation}
m = g
\end{equation}
Hence our problem reduces to solving the Klein-Gordon equation
for a scalar field of mass $m$, sourced by the ``electric field'',
$E_1=F_{01}=Q$, within the capacitor.
The initial condition at $t=0$ is given by the requirement that
no fermions be present, or in terms of the bosonic variables,
\begin{equation}
\phi (t=0 ,x) =0 = {\dot \phi} (t=0,x)
\label{initialconditionsI}
\end{equation}

Before we evolve the equations, however, it is interesting to
find static solutions into which the system can evolve
asymptotically.

\subsection{Static Solution}
\label{staticsolution}

In the asymptotic future, $t \to \infty$, we expect the $\phi$
solution to be simply the static solution to (\ref{kleingordoneqn}).
Since the Klein-Gordon equation (\ref{kleingordoneqn}) is linear,
we may first solve it with
$\bar{F} = (Q/2) (\Theta(x)-\Theta(-x)) = (Q/2){\rm sgn}(x)$
{\it i.e.} due to a single point charge at the origin. The static
solution to the present problem would then follow using appropriate
linear superposition. For now,
\begin{equation}
(\partial^2 + m^2) \phi = - m \frac{Q}{2} \text{sgn}(x)
\label{kleingordoneqn_OneQ}
\end{equation}
Using the integral representation of the step-function and that of
the retarded Green function, $G_{\text{r}}(x-y)$,
\begin{equation}
\Theta(x) = \int \frac{dk}{2 \pi i} \frac{e^{i k x}}{k - i 0^+},
\end{equation}
\begin{equation}
G_{\text{r}}(x-y) = \int \frac{d^2 k}{(2\pi)^2}
          \frac{e^{ik\cdot (x-y)}}{-(k_0-i0^+)^2 + (k_1)^2 + m^2}
\end{equation}
provides us with the integral representation
\begin{eqnarray}
\phi(x) &=& -m \frac{Q}{2} \int d^2y ~ G_{\text{r}}(x-y)
                    (\Theta(y)-\Theta(-y)) \nonumber \\
    &=& -m Q \int \frac{dk}{2\pi} \frac{\sin(kx)}{k(k^2 + m^2)}
\label{staticintegralrep}
\end{eqnarray}

The integral in (\ref{staticintegralrep}) may be evaluated by
performing a partial fraction decomposition of the denominator
$k(k^2+m^2)$ and converting the resulting three sub-integrals into
appropriate contour integrals, which may then be computed
straightforwardly. The answer is
\begin{equation}
\phi_0 (x) = -\frac{Q}{2m} \text{sgn}(x) \left( 1-e^{-m|x|} \right)
\label{staticsolution_OneQ}
\end{equation}
The static solution to (\ref{kleingordoneqn}) is therefore
\begin{eqnarray}
\phi_{\text{s}}(x) &=& - \phi_0(x-L/2) + \phi_0(x+L/2) \nonumber \\
&& \hskip -2 cm = -\frac{Q}{m} \times \left\{
\begin{array}{ll}
e^{-m|x|} \sinh(mL/2),    & |x| > L/2 \\
1-e^{-mL/2} \cosh(mx),    & |x| < L/2
\end{array} \right.
\label{staticsolution_E+E-}
\end{eqnarray}

\subsection{Dynamical Solution}
\label{setup1dynamical}

The solution to (\ref{kleingordoneqn}) we are seeking must
satisfy the initial conditions in (\ref{initialconditionsI}).
To obtain this dynamical solution, we add a homogeneous
solution, $\phi_{\text{h}}$, obeying
$(\partial^2 + m^2)\phi_{\text{h}} = 0$, to the static solution
$\phi_{\text{s}}$ such that the initial conditions are satisfied.
Again, it helps to first solve the problem with a single charge.
Then we have to solve (\ref{kleingordoneqn_OneQ}) with the
initial conditions corresponding to (\ref{initialconditionsI}).
From the conditions in (\ref{initialconditionsI}), we observe
that the integral representation of the solution is
\begin{eqnarray}
\bar{\phi}(t,x) &=& \phi_{\text{s}}(x) + \phi_{\text{h}}(t,x)
 \nonumber \\
&=& -m Q \int \frac{dk}{2\pi} \frac{\sin(kx)}{k(k^2 + m^2)} \nonumber \\
&& \hskip 1 cm \times \left( 1 - \cos\left( t \sqrt{k^2+m^2} \right) \right)
\end{eqnarray}

The solution to (\ref{kleingordoneqn}) with the same initial conditions
is thus
\begin{eqnarray}
\phi(t,x) &=& -\bar{\phi}(t,x-L/2) + \bar{\phi}(t,x+L/2) \nonumber \\
&=& -2 m Q \int \frac{dk}{2\pi} \frac{\cos(kx) \sin(kL/2)}{k(k^2 + m^2)}
                   \nonumber \\
&& \hskip 1 cm \times \left( 1 - \cos\left( t \sqrt{k^2+m^2} \right) \right)
\label{PPphisolution}
\end{eqnarray}
From this, we can extract the flux of energy passing through a given
spatial point $x$ and integrate over all time to get the total
energy radiated. It is
\begin{eqnarray}
\mathcal{F}(x) &=& \int_0^\infty dt \ T^{01} =
     -\int_0^\infty dt \ \partial_0 \phi \partial_1 \phi \nonumber \\
&=& -(2mQ)^2 \int \frac{dk}{2\pi} \int \frac{dp}{2\pi}
          \frac{\sin(k L/2) \cos(kx)}{k(k^2 + m^2)} \nonumber \\
&& \hskip 2 cm \times \frac{\sin(p L/2) \sin(px)}{k^2-p^2}
\end{eqnarray}
We first use
\begin{equation}
\int_{-\infty}^{+\infty} \frac{dk}{2\pi}
\frac{\exp(ikx)}{k^2-a^2} = -\frac{1}{2} \frac{\sin(a|x|)}{a},
\end{equation}
followed by
\begin{equation}
\int\frac{dk}{2\pi} \frac{\exp(ikx)}{k^2(k^2+a^2)} =
     - \frac{1}{2a^2} \left( |x| + \frac{e^{-a|x|}}{a} \right),
\end{equation}
for $a > 0$, to obtain for $|x| > L/2$,
\begin{eqnarray}
\mathcal{F}(x) &=& \text{sgn}(x) \frac{Q^2}{2m}
 \biggl [ \frac{mL}{2} - \sinh\left( \frac{mL}{2} \right)
                 \nonumber \\
       && \hskip 0 cm
\times \biggl \{ e^{-mL/2}
 - e^{-2m|x|} \sinh\left(\frac{mL}{2}\right) \biggr \} \biggr ],
\label{paralellplatecapacitorflux1}
\end{eqnarray}
and, for $|x| < L/2$,
\begin{equation}
\mathcal{F}(x) = \frac{Q^2}{4m} \left( 2 m x - e^{-mL} \sinh(2mx) \right).
\label{paralellplatecapacitorflux}
\end{equation}

By considering the limit $|x| \to \infty$, we obtain the total energy
that is radiated
\begin{equation}
\mathcal{F}_{\text{rad}} \equiv 2 \mathcal{F}( \infty ) =
\frac{Q^2 L}{2} \left [ 1 - \frac{1}{m L} \left( 1 - e^{-mL}
            \right) \right ]
\end{equation}
We can check that the expressions for $\mathcal{F}(x)$ are consistent
with energy-momentum conservation, $\partial_x T^{01} = -\partial_t T^{00}$,
if the final field configuration is the static $\phi_{\text{s}}$. A direct
calculation verifies
\begin{equation}
\partial_x \mathcal{F}(x) = T^{00}(t=0,x) - T^{00}(t=\infty,x)
\end{equation}
with
\begin{equation}
T^{00} = \frac{1}{2} (\partial_t \phi)^2 + \frac{1}{2} (\partial_x \phi)^2
     + \frac{1}{2} \left( m \phi+\bar{F} \right)^2
\end{equation}
where the ``mass term'' in $T^{00}$ arises from the electric field
energy density $(1/2) F_{01}^2$, since $F_{01} = m \phi + \bar{F}$.

One may also use formula 3.876.1 in Gradshteyn and Ryzhik
\cite{GradshteynRyzhik}
\begin{eqnarray}
\int_0^\infty dk \frac{\sin\left( p \sqrt{k^2+a^2}\right)}{\sqrt{k^2+a^2}}
 \cos(bk) && \nonumber \\
&& \hskip -4 cm
= \left \{
\begin{array}{ll}
\frac{\pi}{2} J_0 \left( a \sqrt{p^2-b^2} \right), & 0<b<p \\
0, &  b > p > 0
\end{array} \right.
\label{integralgandr}
\end{eqnarray}
and apply it to the integral representation \eqref{PPphisolution},
to obtain the rate of charge creation,
$\partial_t j^0 \propto \partial_t \partial_x \phi$:
\begin{eqnarray}
\partial_t \partial_x \phi(t,x) &=& \frac{m Q}{2} \biggl [
    J_0\biggl (m \sqrt{t^2- x_-^2 } \biggr ) \Theta (t-x_-)
             \nonumber \\
    &-& J_0\biggl (m \sqrt{t^2- x_+^2 }\biggr )
        \Theta (t-x_+) \biggr ]
   \label{+E-E_chargecreationrate}
\end{eqnarray}
where $x_\pm \equiv x \pm L/2$.
The first term can be attributed to the $-Q$ charge at
$x = +L/2$ whereas the second to the $+Q$ charge at $x = -L/2$.
Charge creation at any location $x$ goes to zero at late times
because the Bessel functions tend to $J_0(m|t|)$ and hence
$\partial_t j^0 \to 0$. This is consistent with the expectation
that the asymptotic field configuration is $\phi_{\text{s}}(x)$
in \eqref{staticsolution_E+E-}.

The electric current at any spacetime location is
\begin{equation}
j^x = g {\dot \phi} =
-2 m^2 Q \int \frac{dk}{2\pi}
     \frac{\cos(kx)\sin(kL/2)}{k \omega_k} \sin(\omega_k t)
\end{equation}
where we have introduced $\omega_k = \sqrt{k^2+m^2}$ and also
used $g=m$. We use a trick to evaluate this integral. Let us
first differentiate with respect to $l\equiv L/2$. Then after
applying appropriate trigonometric identities and
Eq.~(\ref{integralgandr}), this gives
\begin{eqnarray}
\partial_l j^x &=&
- \frac{2m^2 Q}{\pi} \int_0^\infty dk
 \frac{\cos(kx)\cos(kl)}{\omega_k} \sin(\omega_k t) \nonumber \\
&=&
- \frac{m^2 Q}{2} \biggl [
J_0 \biggl ( m\sqrt{t^2-x_-^2} \biggr ) \Theta(t-x_-) \nonumber \\
 && \hskip 1 cm
 + J_0 \biggl ( m\sqrt{t^2-x_+^2} \biggr ) \Theta(t-x_+) \biggr ]
\end{eqnarray}
where we have defined $x_\pm = x\pm l$. Noting that the current
vanishes when the plate separation vanishes ($l=0$), we get
\begin{eqnarray}
j^x (t,x) &=&
- \frac{m^2 Q}{2} \int_0^{L/2} dl \biggl [
J_0 \biggl ( m\sqrt{t^2-x_-^2} \biggr ) \Theta(t-x_-) \nonumber \\
&& \hskip 1 cm
+ J_0 \biggl ( m\sqrt{t^2-x_+^2} \biggr ) \Theta(t-x_+) \biggr ]
\end{eqnarray}

At late times, $t \gg |x_\pm|$, we can Taylor expand the Bessel
functions at $mt$ and then perform the integration over $l$
to get
\begin{eqnarray}
j^x (t,x) &=& -\frac{Q m^2L}{2} J_0(mt) +
 {\mathcal O}\biggl ( \frac{ Q m^3  L^3}{t} J_0'(mt) \biggr ) \nonumber \\
&=& -\frac{Q m^2 L}{\sqrt{2\pi m t}}
       \cos \left (mt-\frac{\pi}{4}\right ) +
 {\mathcal O}\biggl ( \frac{ Q m^3 L^3}{t\sqrt{mt}} \biggr ) \nonumber \\
\label{jxresult}
\end{eqnarray}
where we have used the asymptotic form of the Bessel function
\cite{GradshteynRyzhik} in the second line. The first term is
a good approximation for $t \gg mL^2, ~ |x_\pm|, ~ m^{-1}$.

The expression in Eq.~(\ref{jxresult}) shows that the current
within the capacitor (say at $x=0$) oscillates at the
``microscopic'' frequency given by $m$. If we average out
these fast oscillations, the cosine gets replaced by
$1/\sqrt{2}$ and we find that the root-mean-squared
current decays as $t^{-1/2}$:
\begin{equation}
j_{\rm rms}^x = \frac{m^2QL}{2\sqrt{\pi m t}}
\end{equation}

The electric field within the capacitor decays to a static value
that can be obtained from the static solution
Eq.~(\ref{staticsolution_E+E-}) inserted into (\ref{electricfield}).
The time-dependent electric field within the capacitor at late
times can be obtained from the expression for the current in
Eq.~(\ref{jxresult}) together with the asymptotic static solution
\begin{equation}
E(t,x) = E_{\rm static} -\frac{mQL}{\sqrt{2\pi m t}}
       \sin \left (mt-\frac{\pi}{4}\right )
\label{Eresult}
\end{equation}
To check this expression simply differentiate with respect to
time and keep the leading order term in $1/(mt)$. This agrees
with ${\dot E} = g{\dot \phi} = j^x$. Note that the
static part of the electric field plays no role. In fact,
well away from the capacitor plates, the static electric
field dies off exponentially fast and can be ignored. So we
will define the decaying part of the electric field as
$E_{\rm d} \equiv E-E_{\rm static}$ and refer to this as the
electric field.

The expression for the electric field shows that it is $90^\circ$
out of phase with the current but the amplitude has the same
$1/\sqrt{t}$ decay as the current. The rms value of the electric
field is
\begin{equation}
E_{\rm d,rms} = \frac{mQL}{2\sqrt{\pi m t}}
\end{equation}
This leads to Ohm's law
\begin{equation}
j_{\rm rms}^x = \sigma_{\rm E} E_{\rm d,rms}
\end{equation}
where $\sigma_{\rm E}$ is the electrical conductivity of the vacuum
\begin{equation}
\sigma_{\rm E} = g = \frac{e}{\sqrt{\pi}}
\label{sigmasetup1}
\end{equation}
This result is independent of $Q$ and $L$.

Let us now consider the energy in the capacitor. At late
times, the fields approach the static solution whose
energy can be computed using (\ref{staticsolution_E+E-})
\begin{equation}
\int_{-L/2}^{L/2}T^{00} dx = \frac{Q^2}{4m}\left( 1 - e^{-2Lm} \right).
\label{expectedE}
\end{equation}
This shows that the final energy is smaller if the coupling $g=m$
is stronger. Or stronger coupling implies more complete radiation
of the capacitor energy.

We can identify a typical time scale for energy loss from the
capacitor by considering the ratio of the decaying part of the
energy in the electric field within the capacitor at time $t$
to the initial energy ($Q^2L/2$). The ratio is
\begin{equation}
\frac{{\cal E} (t)}{{\cal E}(0)} \approx
     \frac{m^2 Q^2 L^3/(8\pi m t)}{Q^2 L/2} \equiv \frac{\tau}{t}
\end{equation}
where
\begin{equation}
\tau = \frac{gL^2}{4\pi}
\label{tauvsg}
\end{equation}
Hence larger couplings imply longer decay times {\it i.e.} slower
decay.  The capacitor is more effectively discharged when the
coupling constant is large but it takes a longer time for the discharge
to happen. In the zero coupling limit, the rate of pair production is
rapid, but the original electric field $E$ remains relatively
undissipated.

\section{Setup II: External Plates}
\label{setupII}

\begin{figure}
\includegraphics[width=3.0in]{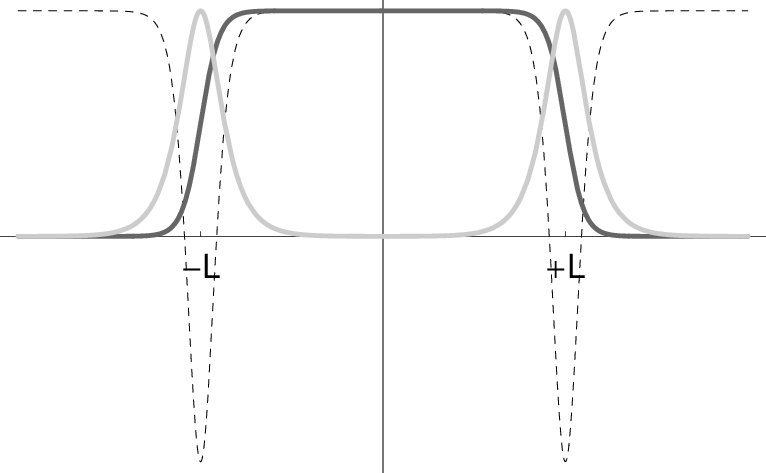}
\caption{
Schematic view of setup II. The dashed line is the double
well potential $V(x+L)+V(x-L)$. The thick black line is the initial
$\phi$ configuration, $\tanh(x+L)-\tanh(x-L)$, and the thick grey
line is a bound state solution in the double well. Note that the
figure is meant to be schematic and the horizontal axis is not the
zero of the potential.
}
\label{DoubleWell_Setup}
\end{figure}

In the second setup we do not wish to introduce external charges.
Instead the capacitor is charged with the same fermionic
field, $\psi$, or its bosonized version, $\phi$.
However we still need to
have some capacitor ``plates'' that we can charge. These
plates have to be external to the system. To implement this
scheme, we add a double well potential to the action in
(\ref{bosonicaction})
\begin{equation}
\mathcal{S}_{\text{V}} \equiv
-\frac{1}{2} \int d^2 x [V(x+L)+V(x-L)] \phi^2 \\
\label{doublewellaction}
\end{equation}
where the form of $V(x)$ is chosen so that we can find a
non-dissipative solution for the scalar field in a single
well. A sketch of the setup is shown in Fig.~\ref{DoubleWell_Setup}.\footnote{See \cite{Bazeia:2002xt} for a setup in a similar spirit, but in the context of QCD$_2$.}

It is convenient to choose
\begin{equation}
V(x) = - 2 M^2 \text{sech}^2(M x)
\label{doublewellform}
\end{equation}
where $M$ is some mass scale that sets the depth and width of
the well. With this choice the solution to the single well problem
\begin{equation}
(\partial^2 + m^2 + V(x)) \phi = 0
\end{equation}
is given by the $\lambda\phi^4$ kink
\begin{equation}
\phi(t,x) = \cos(m t) \tanh(Mx) ,
\end{equation}
where we have chosen the initial condition ${\dot \phi}(t=0,x) = 0$.
The solution describes a positive charge in the well at $t=0$, which
then oscillates -- due to pair creation in the electric field --
but does not dissipate. The charge in the well at any given time
can be found from Eq.~(\ref{maxwelleqn}) to be $2m\cos(mt)$.

Now consider a capacitor with two plates, one at $x=L$ and the other
at $x=-L$. Since we do not have any external charges, we take the
constant background electric field, ${\bar F}$ in
Eq.~(\ref{electricfield}), to be zero. This means we are now solving
\begin{equation}
(\partial^2 + m^2 + V(x+L) + V(x-L)) \phi = 0
\label{doublewelleqn}
\end{equation}
with the following choice of initial conditions
\begin{eqnarray}
\phi(t=0,x) &=& \tanh(M(x+L)) - \tanh(M(x-L)) \equiv \phi_0 (x)
                  \nonumber \\
 {\dot \phi}(t=0,x) &=& 0
\label{phi0}
\end{eqnarray}
In what follows, we shall set $M=1$ and so all quantities will
be in units of $M$.

\subsection{Asymptotic State}
\label{setupIIasymptotic}

Before solving for the time evolution, we consider the asymptotic state,
which will be a stationary solution of Eq.~(\ref{doublewelleqn}).
That is, we think of the double well equation (\ref{doublewelleqn})
as a Schrodinger equation
\begin{equation}
H \psi_n = \omega_n^2 \psi_n
\end{equation}
with Hamiltonian
\begin{equation}
H \equiv -\partial_x^2 + m^2 + V(x+L) + V(x-L)
\label{hamiltonian}
\end{equation}

The corresponding Hamiltonian for a single potential well centered
at $x=0$ is
\begin{equation}
H_1 = -\partial_x^2 + m^2 + V(x)
\end{equation}
This has exactly one bound state
\begin{equation}
\psi \propto \text{sech}(x)
\end{equation}
with eigenvalue $\omega_0^2 = m^2-1$ \cite{morsefeshbach}.
(Recall that we are working in units with $M=1$.)
So for the double well potential, at least when the two wells are
well separated, there must be exactly two bound states, which
can be approximated as
\begin{equation}
\psi \sim {\rm sech}(x-L) \pm {\rm sech}(x+L)
\label{boundstate2}
\end{equation}
The energies of these two bound states are nearly identical,
$\omega^2_\pm = m^2-1 \pm e^{-\Gamma}$, $\Gamma(m,L) \gg 1$, split
by exponentially small corrections due to tunneling between the double
wells. These bound states are the stationary states that the system
can evolve into.

We also observe that there is an apparent instability in the current
model when $m^2<1$, since then $\omega_0^2 < 0$ and the bound state
solution can grow exponentially. To understand this instability, we
examine the double well action in (\ref{doublewellaction}). If the
potential $V(x)$ is deep enough, there will be a region where
$m^2 + V(x)$ is sufficiently negative, that it becomes favorable
for $\phi$ to grow without bound in this region. In terms of the
fermionic model, the well is so deep that it is favorable to pull
fermion pairs out of the vacuum and put them at the bottom of the
potential.

The evolution of the initial data, $\phi_0$, in this setup can be
evolved formally by writing
\begin{equation}
\phi(t,x) =
 \sum_n \cos\left(\omega_n t\right) \psi_n(x) \langle \psi_n | \phi_0 \rangle,
\label{doublewellsolution_expansion}
\end{equation}
where the summation is over both bound and continuum states of $H$.
We can check that the initial conditions are satisfied by setting
$t=0$ in the factor $\cos(\omega_n t)$ and in its time derivative.
We expect that, as time progresses, the continuum states will disperse
to infinity, leaving behind only the initial overlap with the bound
state. While formally correct, the expansion in
Eq.~(\ref{doublewellsolution_expansion}) is only useful if we know
the full eigenspectrum of the double well potential. In the absence
of the eigenspectrum, it is easier to numerically evolve the equation
of motion.

\subsection{Time Evolution}
\label{setupIItimeevolution}

We have evolved Eq.~(\ref{doublewelleqn}) using the explicit
Crank-Nicholson algorithm with two iterations with first-order
absorbing boundary conditions. The runs were done on very large
lattices so that boundary effects are minimal.

\begin{figure}
\includegraphics[width=3.4in]{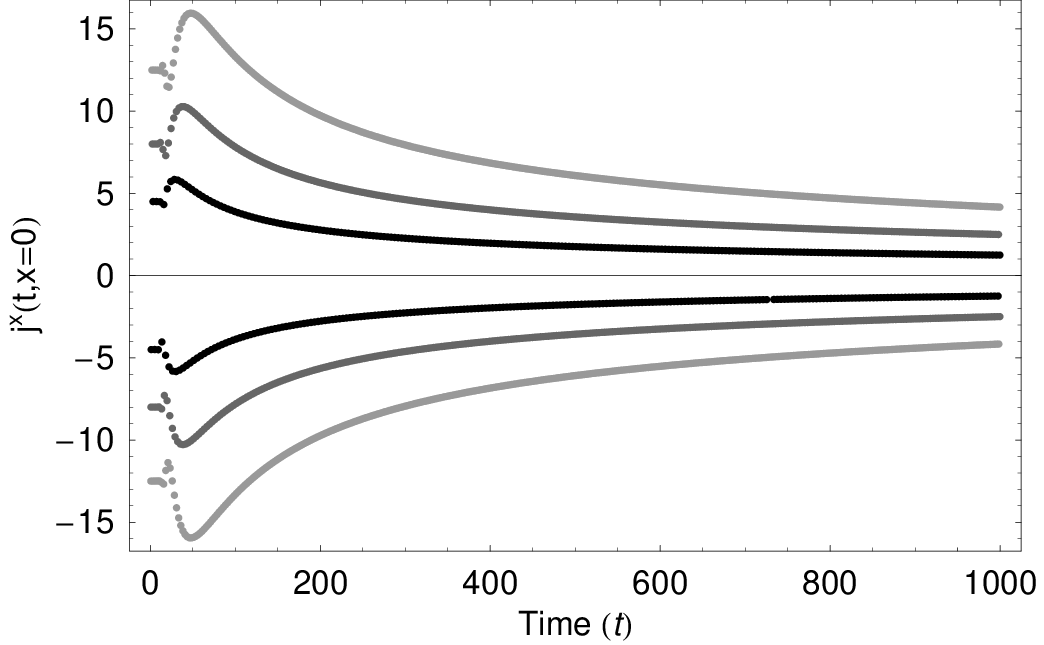}
\caption{
Envelopes of the plots of the current at the center of the capacitor,
$j^x(t,x=0)$, versus time. From black to light grey, the curves
represent, respectively, the evolution for $m = 1.5, 2$ and 2.5.
The rapid oscillations between the envelopes are not shown.
}
\label{CurrentDensity_DoubleWell}
\end{figure}

\begin{figure}
\includegraphics[width=3.4in]{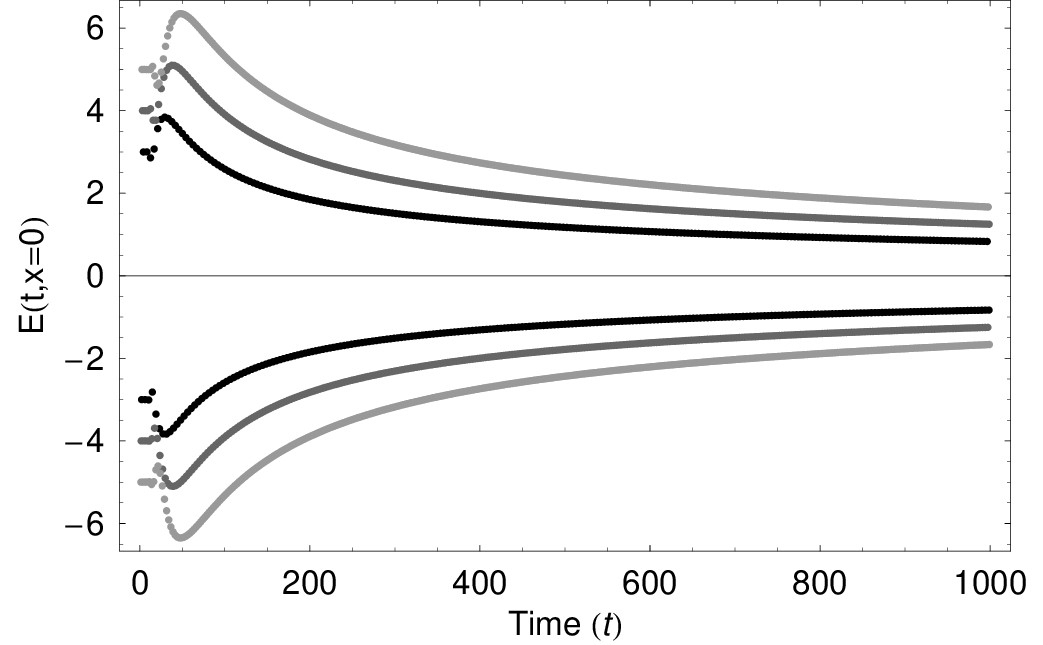}
\caption{
Envelopes of the plots of the electric field at the center of the
capacitor, $E(t,x=0)$ versus time. From black to light grey, the curves
represent, respectively, the evolution for $m = 1.5, 2$ and 2.5.
The rapid oscillations between the envelopes are not shown.
}
\label{ElectricField_DoubleWell}
\end{figure}

\begin{figure}
\includegraphics[width=3.4in]{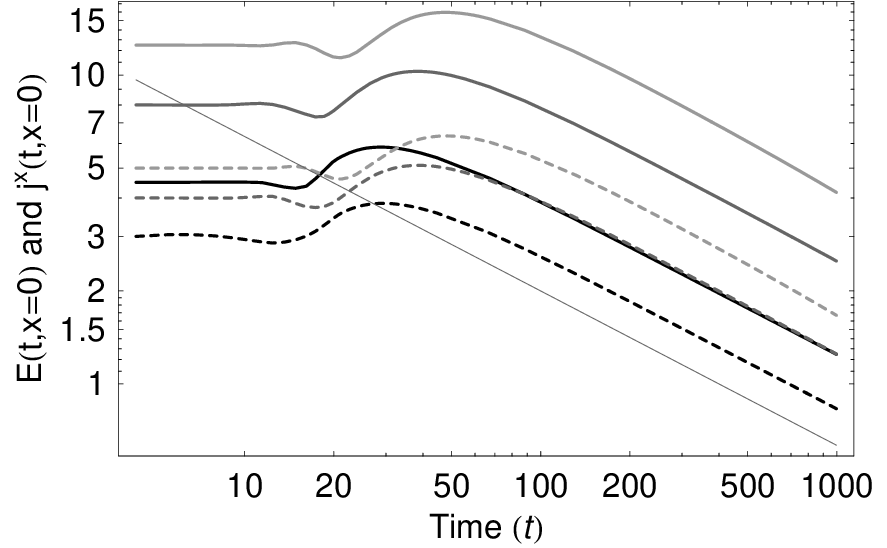} \\
\caption{Envelopes for the currents (solid curves) and the
electric fields (dashed curves) at $x=0$ on a log-log plot. Comparison
with the light straight line with slope $-1/2$ clearly shows the
$1/\sqrt{t}$ fall off. From black to light grey, the curves
represent, respectively, the evolution for $m =1.5$, 2 and 2.5.
}
\label{JandELogLog}
\end{figure}

In Fig.~\ref{CurrentDensity_DoubleWell} we plot the current envelopes
at $x=0$ versus time for several different parameters, disregarding
the rapid oscillations between the envelopes. Similarly, in
Fig.~\ref{ElectricField_DoubleWell} we show the behavior of the
electric field at $x=0$. On the log-log plot in
Fig.~\ref{JandELogLog}, it is clear that the envelopes
decay as a power law. A fit gives
\begin{equation}
E_{\rm rms} =  N \sqrt{\frac{g}{t}} \ , \ \
j^x_{\rm rms} = g N \sqrt{\frac{g}{t}}
\end{equation}
where $N \approx 14$ is a factor which could depend on the
dimensionless product $ML$, where $M^{-1}$ is the width
of the wells (see Eq.~(\ref{doublewellform})). The electrical
conductivity is therefore again given by Eq.~(\ref{sigmasetup1}),
as for setup 1.

\begin{figure}
\includegraphics[width=3.4in]{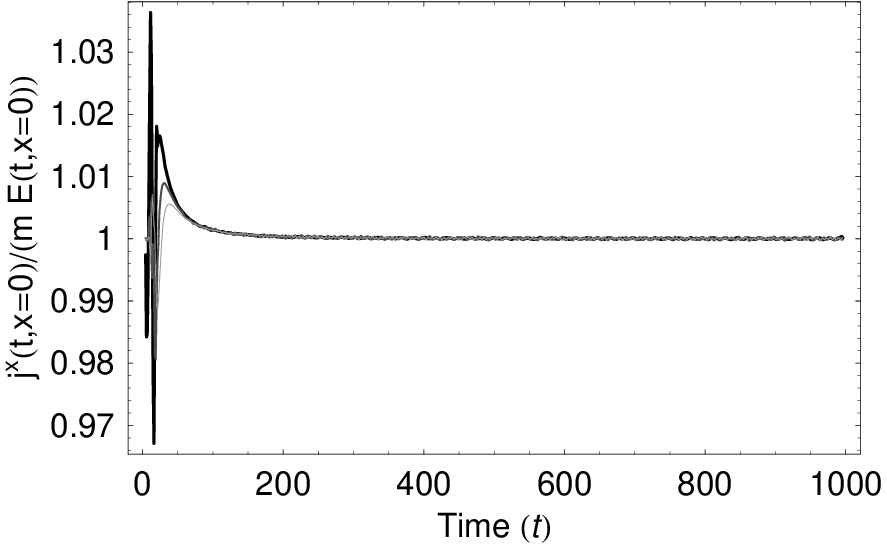}
\caption{Ratio of the envelopes of current (divided by $m$) to electric
field at $x=0$ for $m=1.5$, 2 and 2.5. The constant flat ratio implies
Ohm's law and the value of 1 implies that the conductivity is $g$.
}
\label{Conductivity_DoubleWell}
\end{figure}

\section{Conclusions and Discussion}
\label{conclusions}

We have studied the quantum discharge of a capacitor in massless
QED in 1+1 dimensions by bosonizing the model. The bosonized model
is non-interacting and can be solved classically. The solution
includes all backreaction effects. We now summarize the key results.

The final state depends on the setup used to describe the
capacitor plates. We have chosen two different ways to describe
the capacitor plates. In both cases, the final state is non-trivial.
In setup I, the plates keep their original charge but the charges
are screened due to the Schwinger process. In setup II, there are
no external charges, but there are external potentials that play
the role of capacitor plates. Then the final state consists of
fermion-antifermion pairs that are bound to the plates.

The energy in the final state depends on the coupling constant,
$g$, and equivalently the mass of the scalar field, $m$. The final
state energy decreases with increasing $g$, while the time for the
capacitor to discharge increases with increasing $g$, as seen
in Eq.~(\ref{tauvsg}). So stronger coupling leads to more complete
discharge but the discharge process itself is slower. We
suggest that the longer discharge time for larger coupling
constant is due to the tighter binding of fermion-antifermion
pairs that need to be split apart by the electric field.

The discharge process is highly oscillatory, as also seen in the
semiclassical analysis \cite{CooperMottolaEtal} and the amplitude
of oscillations falls off rather slowly, as $t^{-1/2}$. This
suggests that the massless QED system is under damped

Our results show that the root-mean-square current in the capacitor
is directly proportional to the root-mean-square electric field,
indicating that Ohm's law holds on a macroscopic scale. Thus it
makes sense to define
the electrical conductivity for the massless QED vacuum to be
$\sigma_{\rm E} = j_{\rm rms}^x/E_{\rm rms}$ and our
results indicate the simple relation $\sigma_{\rm E} = g$ which
can also be written in terms of the fermionic charge as
$\sigma_{\rm E}= e/\sqrt{\pi}$.

A correspondence is often made between Schwinger pair creation
and Hawking radiation, though we have indicated differences between
the two processes that prompt us to use caution in drawing a
correspondence. If the oscillatory features of the discharge
process carry over to black hole evaporation, we may expect
black hole mass oscillations during evaporation. Though, in
contrast to the capacitor, the black hole system is unstable
in that smaller mass black holes are hotter and evaporate faster,
while weaker electric fields in the capacitor do not discharge faster.
So it would appear that a fluctuation that excessively decreases
the mass of the black hole, would make it evaporate yet faster
in what may be a runaway process.

The issues of black hole formation and the final state of black
hole evaporation cannot be resolved by this correspondence since
the capacitor plates have to be introduced externally, whereas
there are no such externally set conditions in the case of
gravitational collapse. Yet it would be extremely interesting
if the electromagnetic Ohm's law has a gravitational analog that
relates energy flow (current) from a black hole, or during
gravitational collapse, to the ``gravitational electric'' field
(see Sec.~4.4 of \cite{Waldbook}). Perhaps the instability of
the black hole can be summarized in a negative ``gravitational
conductivity''.

A potential application of our findings is to superconducting cosmic
strings, where massless QED in 1+1 dimensions is expected
to apply for fermion zero modes on the string  \cite{Witten:1984eb}.
Our analysis shows that if superconducting strings really behave as
1+1 dimensional systems, they will carry oscillatory currents because
of the backreaction of the induced currents on the external electric
fields. (Oscillatory currents were also discussed in \cite{Aryal:1987pw},
though these occurred due to the periodic dynamics of the strings.)

While our analysis has enabled us to fully treat backreaction
of the Schwinger process, our results cannot be transported to
3+1 QED for two reasons. First, the electron has a non-zero mass.
For electric fields smaller than the electron mass squared, the
Schwinger process is exponentially suppressed and the vacuum is
essentially an insulator. In situations where the electric field
is larger than the electron mass squared, the exponential suppression
is absent and the dynamics may be closer to what we have found.
The second reason is that the larger number of dimensions can
change the picture dramatically. In 1+1 dimensions, the inter-charge
potential is linear and electric charge is confined. In 3+1 dimensions,
electric charges interact by the Coulomb potential and are not confined.
This suggests that our system may be closer to the case of
chromo-electric fields in 3+1 dimensions with massless quarks. While
this has some features that resemble the model we have considered,
there are essential differences due to the non-Abelian nature of
the interactions.

\begin{acknowledgments}
We thank Edward Witten for suggesting the massless Schwinger model
for the backreaction problem, Ioannis M. Besieris for sharing his
notes on the solution to the massive Klein-Gordon equation, and
Ratin Akhoury, Daniel Green, Zohar Komargodski, Juan Maldacena, Dmitry I. Podolsky, David Shih and Yuji Tachikawa for helpful comments and discussions.
We thank Peter J. Kernan and Pascal M. Vaudrevange for invaluable
computing help. This work was supported by the U.S. Department
of Energy at Case Western Reserve University. TV was also supported
by grant number DE-FG02-90ER40542 at the Institute for Advanced Study.
\end{acknowledgments}


\end{document}